\documentclass[11pt]{article}
\pdfoutput=1
\usepackage[centertags]{amsmath}
\usepackage[square, comma, sort&compress,numbers]{natbib}
\usepackage{array,multirow}
\numberwithin{equation}{section}
\usepackage{amssymb,amsfonts}
\usepackage{graphicx}
\usepackage{color}
\usepackage{mathtools}

\usepackage{epsfig}
\usepackage{bbold}

\usepackage{wrapfig}
\usepackage{float}
\usepackage{soul}
\usepackage{tkz-euclide}

\usepackage{tikz,pgf}
\usetikzlibrary{shapes}
\usetikzlibrary{calc}
\usetikzlibrary{decorations.pathmorphing}
\usetikzlibrary{decorations.pathreplacing,shapes.misc}
\usetikzlibrary{positioning}
\usetikzlibrary{arrows}
\usetikzlibrary{decorations.markings}
\usetikzlibrary{shadings}

\usetikzlibrary{intersections}

\def\be{\begin{equation}}
\def\ee{\end{equation}}
\def\ba{\begin{array}}
\def\ea{\end{array}}

\def\dps{\displaystyle}
\newcommand{\half}{\frac{1}{2}}

\def\1{\tilde{1}}
\def\2{\tilde{2}}
\def\3{\tilde{3}}

\newdimen\tableauside\tableauside=1.0ex
\newdimen\tableaurule\tableaurule=0.4pt
\newdimen\tableaustep
\def\phantomhrule#1{\hbox{\vbox to0pt{\hrule height\tableaurule
width#1\vss}}}
\def\phantomvrule#1{\vbox{\hbox to0pt{\vrule width\tableaurule
height#1\hss}}}
\def\sqr{\vbox{%
  \phantomhrule\tableaustep

\hbox{\phantomvrule\tableaustep\kern\tableaustep\phantomvrule\tableaustep}%
  \hbox{\vbox{\phantomhrule\tableauside}\kern-\tableaurule}}}
\def\squares#1{\hbox{\count0=#1\noindent\loop\sqr
  \advance\count0 by-1 \ifnum\count0>0\repeat}}
\def\tableau#1{\vcenter{\offinterlineskip
  \tableaustep=\tableauside\advance\tableaustep by-\tableaurule
  \kern\normallineskip\hbox
    {\kern\normallineskip\vbox
      {\gettableau#1 0 }%
     \kern\normallineskip\kern\tableaurule}%
  \kern\normallineskip\kern\tableaurule}}
\def\gettableau#1 {\ifnum#1=0\let\next=\null\else
  \squares{#1}\let\next=\gettableau\fi\next}

\newcommand{\bref}[1]{\textbf{\ref{#1}}}

\def\cF{\mathcal{F}}

\def\cO{\mathcal{O}}

\numberwithin{equation}{section} \makeatletter
\@addtoreset{equation}{section}

\def\be{\begin{equation}}
\def\ee{\end{equation}}
\def\ba{\begin{array}}
\def\ea{\end{array}}
\def\dps{\displaystyle}

\def\ba{\begin{array}}
\def\ea{\end{array}}

\newmuskip\pFqmuskip
\newcommand*\pFq[6][8]{%
  \begingroup 
  \pFqmuskip=#1mu\relax
  \mathcode`\,=\string"8000
  \begingroup\lccode`\~=`\,
  \lowercase{\endgroup\let~}\pFqcomma
  {}_{#2}F_{#3}{\left[\genfrac..{0pt}{}{#4}{#5};#6\right]}%
  \endgroup
}
\newcommand{\pFqcomma}{\mskip\pFqmuskip}

\newmuskip\LpFqmuskip
\newcommand*\LpFq[6][8]{%
  \begingroup 
  \pFqmuskip=#1mu\relax
  \mathcode`\,=\string"8000
  \begingroup\lccode`\~=`\,
  \lowercase{\endgroup\let~}\pFqcomma
  {}_{}F_{}{\left[\genfrac..{0pt}{}{#4}{#5};#6\right]}%
  \endgroup
}

\newmuskip\Ftmuskip
\newcommand*\Ft[6][8]{%
  \begingroup 
  \pFqmuskip=#1mu\relax
  \mathcode`\,=\string"8000
  \begingroup\lccode`\~=`\,
  \lowercase{\endgroup\let~}\pFqcomma
  F_{2}{\left[\genfrac..{0pt}{}{#4}{#5};#6\right]}%
  \endgroup
}
\newmuskip\Ftmuskip
\newcommand*\FK[6][8]{%
  \begingroup 
  \pFqmuskip=#1mu\relax
  \mathcode`\,=\string"8000
  \begingroup\lccode`\~=`\,
  \lowercase{\endgroup\let~}\pFqcomma
  F_{K}{\left[\genfrac..{0pt}{}{#4}{#5};#6\right]}%
  \endgroup
}

\usepackage{jheppub}
\makeatletter
\def\@fpheader{\vspace{-.1cm}}
\makeatother

\title{Differential equations for classical Virasoro blocks with heavy and light operators}

\author[a]{Mikhail\ Pavlov}

\affiliation[a]{I.E. Tamm Department of Theoretical Physics, \\P.N. Lebedev Physical
Institute,\\ Leninsky ave. 53, 119991 Moscow, Russia}

\emailAdd{pavlov@lpi.ru}

\abstract{In this note we study differential equations for classical blocks with heavy and light operators. We present ODEs for the $4$-pt blocks, generalizing the ODE for the $4$-pt identity block, found by Fitzpatrick, Kaplan, Walters and Wang in \cite{Fitzpatrick:2015foa}.}

\makeatletter
\def\@fpheader{\vspace{-.1cm}}
\makeatother

\begin{document}

\maketitle 
\flushbottom

\flushbottom

\section{Introduction}

The main objects in two-dimensional conformal field theories (CFT$_2$) are correlation functions of primary operators which can be expanded in {\it conformal blocks} \cite{Belavin:1984vu, Zamolodchikov1986}. Conformal blocks represent basis in the space of correlation function which are determined only by the Virasoro algebra. They are usually known only in the form of series \cite{Hadasz:2005gk}, but or the exceptional cases like special values of the Virasoro algebra's central charge, conformal blocks are related to non-linear differential equations  (for example\footnote{Also, the case for $c=-2$ was studied in \cite{Bershtein:2018zcz}.}, through the Kiev's formula for $c=1$ \cite{Gamayun:2012ma}). Another direction is to consider various ($c \to \infty$)  classical limits \cite{Hadasz:2005gk, Fateev:2011qa,  Litvinov:2013sxa, Harlow:2011ny, Besken:2019jyw}.   

The essential step forward is the approximation of heavy and light operators (the HL approximation) \cite{Hartman:2013mia, Fitzpatrick:2014vua}. We will refer to the classical conformal blocks with heavy and light operators as to {\it HL blocks} beneath. The HL approximation not only allows to find HL blocks in many cases explicitly \cite{Fitzpatrick:2014vua, Hijano:2015rla, Fitzpatrick:2015zha, Alkalaev:2015lca, Alkalaev:2018nik}, but also provides the clear AdS/CFT description of such blocks in terms of geodesic graphs \cite{Hartman:2013mia, Alkalaev:2015wia} or Steiner trees \cite{Alkalaev:2018nik}. 

The interesting property of HL blocks was revealed in \cite{Fitzpatrick:2015foa}, where it was pointed out that a classical $4$-pt  identity block with two heavy and two light operators satisfies a Riccati equation.\footnote{See \cite{Karlsson:2021mgg} for the discussion related to W$_N$ algebras.}  This result stems from a specialized diagrammatic technique for classical blocks developed in that work. Crucially, analogous structures appear in the stress-tensor sector of 4-pt blocks for $d=4,6,8$ \cite{Kulaxizi:2018dxo, Karlsson:2021mgg}. However, despite the technique's utility, its generalization beyond 4-pt HL identity blocks remains unknown.  Motivated by this limitation, the main goal of this paper is to present a way to derive equations on HL blocks without the diagrammatic technique, but instead using the monodromy method or the dual description. 
 
The paper is organized as follows. In Section \bref{sec:HL_B} we review HL blocks calculations: from the monodromy method standpoint (subsection \bref{sec:HL_C}) and from the AdS/CFT perspective (subsection \bref{sec:HL_A}). Section \bref{sec:MODE} is devoted to derivation of various ODEs for HL blocks from the monodromy method and a relation between such ODes and BPZ equations.  In Section \bref{sec:HODE} we discuss an interplay between the ODEs mentioned above and AdS description of HL blocks focusing on the case of the HHLL blocks. 
Concluding Section \bref{sec:conclusion} summarizes our results and contains future developments.

\section{HL classical blocks: the monodromy method and Steiner trees on the Poincare disk}
\label{sec:HL_B}

This Section recalls the monodromy method for computing HL blocks and their AdS description in terms of Steiner trees on the Poincare disk. We mainly focus on the HL blocks with two and three heavy operators, which are analyzed in Section \bref{sec:MODE}.  

\subsection{CFT$_2$ side}
\label{sec:HL_C}
\paragraph{Classical conformal blocks within the monodromy method.} 

Let $\cF_n(z|\Delta_i, \tilde{\Delta}_p, c)$ denote the holomorphic Virasoro conformal $s$-channel block corresponding to the $n$-point correlation function of primary operators (with conformal dimensions $\Delta_i$) inserted at points $z_i$ for $i = 1, \dots, n$ \cite{Belavin:1984vu, DiFrancesco:1997nk}. The block also depends on intermediate conformal dimensions $\tilde{\Delta}_p$ (for $p = 1, \dots, n-3$) and the central charge $c$ of the Virasoro algebra.

In the classical limit, where $c \to \infty$ while $\Delta_i/c$ and $\tilde{\Delta}_p/c$ remain finite, the block $\cF_n(z|\Delta_i, \tilde{\Delta}_p, c)$ is conjectured to take an exponential form \cite{Zamolodchikov1986, Hadasz:2005gk, Besken:2019jyw}:
\be
\label{class}
\cF_n(z|\Delta_i, \tilde{\Delta}_p, c) = \exp\left[\frac{c}{6} f_n(z|\epsilon_i, \tilde{\epsilon}_p)\right] + \cO\left(\frac{1}{c}\right), \quad \text{as} \quad c \to \infty,
\ee
where $\epsilon_i \equiv 6 \Delta_i / c$ and $\tilde{\epsilon}_p \equiv 6 \tilde{\Delta}_p / c$ are the classical dimensions, and $f_n(z|\epsilon_i, \tilde{\epsilon}_p)$ is the $n$-point classical conformal block.

Within the monodromy method, we introduce an auxiliary block $\Psi(y|z)$ corresponding to an $(n+1)$-point correlation function involving the original $n$ primaries and an additional degenerate operator $V_{(2,1)}(y)$ \cite{Belavin:1984vu}. The degenerate operator satisfies the null-state condition
\be
\label{V21}
\left(L_{-2} - \frac{3}{2(2 \Delta_{\psi} + 1)} L_{-1}^2 \right)|\psi \rangle = 0,
\ee
where $\Delta_{\psi} = -\frac{1}{2} - \frac{3}{4}b^2$, and the parameter $b$ is related to the central charge via $c = 1 + 6(b + b^{-1})^2$. 

The presence of the degenerate operator $V_{(2,1)}(y)$ in the correlation function ensures that the correlator (and hence, the corresponding conformal block) satisfies the BPZ equation. This equation arises from representing the Virasoro generators in the null-state condition \eqref{V21} as differential operators acting on the conformal block. Moreover, since $\Delta_\psi$ remains finite as $c \to \infty$, the auxiliary block factorizes in the classical limit (see Appendix C.2 of \cite{Fitzpatrick:2014vua} for the derivation)
\be
\label{factor}
\Psi(y|z) = \psi(y|z) \exp\left[\frac{c}{6} f_n(z|\epsilon_i, \tilde{\epsilon}_p)\right], \quad c \to \infty.
\ee

The conditions \eqref{factor} and \eqref{V21} in the classical limit are reduced to the following BPZ equation (see Appendix C.3 of \cite{Fitzpatrick:2014vua} for details):
\be
\label{BPZ}
\left[\frac{d^2}{dy^2} + T(y|z)\right] \psi(y|z) = 0,
\ee
where the stress tensor is given by
\be
\label{TG}
T(y|z) = \sum_{i=1}^{n} \left( \frac{\epsilon_i}{(y-z_i)^2} + \frac{c_i}{y-z_i} \right), \qquad c_i = \frac{\partial}{\partial z_i} f_n(z|\epsilon_i, \tilde{\epsilon}_p).
\ee
Here, the accessory parameters $c_i$ encode the dependence of the $n$-point classical block on the insertion points $z_i$. Using global $SL(2)$ invariance, we can fix three insertion points $\{z_1,z_{n-1},z_{n}\}$ to $\{0, 1, \infty\}$, reducing the number of independent accessory parameters to $(n-3)$. More precisely, the stress tensor \eqref{TG}  must decay as $T(y|z) \sim y^{-4}, y \to \infty$, leading to the linear equations for $c_1, c_{n-1}$ and $c_{n}$. One can solve these equation and substitute back to $T(y|z)$ which gives us the refined form of the stress tensor (which we denote $T(y, z_i)$ for clarity)
\be
\label{T}
T(y, z_i) = \sum_{i=1}^{n-1} \frac{\epsilon_i}{(y - z_i)^2} + \sum_{i=2}^{n-2} c_i \frac{z_i (z_i - 1)}{y(y - z_i)(y - 1)} + \frac{\epsilon_n}{y(y - 1)} - \sum_{i=1}^{n-1} \frac{\epsilon_i}{y(y - 1)}.
\ee

Besides the direct BPZ equation, conformal blocks with the degenerate operator $V_{(2,1)}(y)$ have special monodromy properties \cite{Belavin:1984vu}. Consider a system of concentric contours $\Gamma_k$ encircling subsets of points $\{z_1, \dots, z_{k+1}\}$. As the argument $y$ of the degenerate operator $V_{(2,1)}(y)$ traverses these contours, the solution $\psi(y|z)$ must exhibit prescribed monodromy properties (in the classical limit)
\be
\label{TM}
\widetilde{M}_p = - \begin{pmatrix}
  e^{i \pi  \gamma_p }& 0\\
  0& e^{-i \pi   \gamma_p}
\end{pmatrix}, \qquad \gamma_p = \sqrt{1 - 4 \tilde{\epsilon}_p}, \qquad p = 1,...,n-3. 
\ee

The monodromy method involves solving the BPZ equation \eqref{BPZ} subject to the monodromy conditions \eqref{TM} around the contours $\Gamma_p$. This imposes algebraic constraints on the accessory parameters $c_i$, known as the monodromy equations. Resolving these equations yields a system of first order PDEs, which can then be integrated to determine the classical block $f_n(z|\epsilon_i, \tilde{\epsilon}_p)$.

\paragraph{The HL approximation.}
 The BPZ equation \eqref{BPZ} is a Fuchsian differential equation with $n$ regular singular points at $y = z_i$. While the monodromy around each singular point $z_i$ is fixed, determining the monodromy around contours $\Gamma_p$ (encircling multiple singularities) is challenging for general $n$ since explicit solutions and their monodromy properties are only known for small $n$ (e.g., $n\leq3$). 

 To address this, we employ the heavy-light (HL) approximation \cite{Fitzpatrick:2014vua}, where operators are divided into two subsets: $(n-k)$ heavy operators with large classical dimensions $\epsilon_j$ and $k$ light operators with small classical dimensions $\epsilon_i$ satisfying
\begin{equation}
\label{perl}
\epsilon_i \ll \epsilon_j, \quad i=1,\dots,k, \quad j=k+1,\dots,n.
\end{equation} Hence, in the zeroth order of the HL approximation we end up with the Fuchsian equation with less $(n-k)$ singular points. We focus on cases with two or three heavy operators ($n-k=2,3$). In this approximation, we expand in powers of the light dimensions 
\be
\label{decos}
\begin{gathered}
\psi(y| z) = \psi^{(0)}(y) + \psi^{(1)}(y|z)+ \psi^{(2)}(y|z)+...\,,
\qquad
T(y|z) = T^{(0)}(y) + T^{(1)}(y|z)+T^{(2)}(y|z)...\,,
\\
f( z|\epsilon, \tilde{\epsilon}) = f^{(1)}( z|\epsilon, \tilde{\epsilon})  + f^{(2)}( z|\epsilon, \tilde{\epsilon})  +...\,, \qquad
c_{m}( z|\epsilon, \tilde{\epsilon}) = c_{m}^{(1)}( z|\epsilon, \tilde{\epsilon})  +  c_{m}^{(2)}( z|\epsilon, \tilde{\epsilon})...\,,
 \end{gathered}
\ee
where superscripts denote orders in the light dimensions. Here and below, we fix heavy operator positions at two or three standard points $\{0,1,\infty \}$ (depending on operator count). This eliminates $z_i$-dependence in the zeroth order stress tensor and hence, in $\psi^{(0)}(y)$. Furthermore, the zeroth order classical block reduces to an irrelevant constant, and all accessory parameters vanish.

\paragraph{Two heavy operators.} 
Consider the 4-pt block with two heavy operators $\epsilon_3 = \epsilon_4 = \epsilon_H$ (located at points $1$ and $\infty$), 
two light operators $\epsilon_1, \epsilon_2$ and an intermediate dimension $\tilde{\epsilon}$ (the HHLL block). Under the HL hierarchy, one has
\begin{equation}
\epsilon_1, \epsilon_2, \tilde{\epsilon} \ll \epsilon_H.
\end{equation}
Substituting \eqref{decos} into \eqref{BPZ} and \eqref{T} yields the perturbative system
\be
\label{BPZ012}
\ba{c}
\dps \left(\frac{d^2}{dy^2} + T^{(0)}(y)\right) \psi^{(0)}(y) =0,  ~~~~~~~~~~~~\\
\dps \left(\frac{d^2}{dy^2} + T^{(0)}(y)\right) \psi^{(1)}(y) = - T^{(1)}(y,z) \psi^{(0)}(y),\\
\dps \left(\frac{d^2}{dy^2} + T^{(0)}(y)\right) \psi^{(2)}(y) = - T^{(1)}(y,z) \psi^{(1)}(y) - T^{(2)}(y,z) \psi^{(0)}(y), \\
\ea
\ee
with stress tensor components
\be
\ba{c}
\label{T012}
\dps  T^{(0)}(y)  = \frac{\epsilon_H}{(1-y)^2}\;,
\\
\dps T^{(1)}(y, z) =  \frac{\epsilon_1}{y^2} +  \frac{\epsilon_2}{(y-z)^2}+c^{(1)}_2(z) \frac{ z (z - 1)}{y(y - z) (y - 1)} - \frac{ \epsilon_1 + \epsilon_2}{y (y - 1)}\;, \\
\\
\dps T^{(2)}(y, z) = c^{(2)}_2(z) \frac{ z (z - 1)}{y(y - z) (y - 1)}. 
\ea
\ee
The homogeneous operator in \eqref{BPZ012} remains identical across orders. This allows solving the inhomogeneous equations for $\psi^{(1)}$ and $\psi^{(2)}$ using the variation of parameters method applied to the zeroth order solutions.
 
The zeroth order solutions of \eqref{BPZ012}
are 
\be
\label{2BP}
\psi^{(0)}_{\pm} (y) = \dps (1-y)^{\frac{1\pm \alpha}{2}}, \qquad \alpha = \sqrt{1-4\epsilon_H},
\ee
and the first/second order solutions are
\be
\label{f1s2}
\ba{c}
\dps \psi^{(1)}_{\pm} (y,z) = \frac{1}{W} \left[ \psi^{(0)}_{-} (y) \int^{y} dx ~  \psi^{(0)}_{\pm} (x) T^{(1)}(x,z) \psi^{(0)}_{+} (x) - \psi^{(0)}_{+} (y) \int^{y} dx ~ \psi^{(0)}_{\pm} (x) T^{(1)}(x,z) \psi^{(0)}_{-} (x) \right],  
\ea
\ee
\be
\ba{c}
\label{s3}
\dps \psi^{(2)}_{\pm} (y,z) = \frac{\psi^{(0)}_{-} (y)}{W}   \left[\int^{y} dx ~  (\psi^{(1)}_{\pm} (x,z) T^{(1)}(x,z) + T^{(2)}(x,z)\psi^{(0)}_{\pm}(x))  \psi^{(0)}_{+} (x) \right] - \\
\\
 \dps \frac{\psi^{(0)}_{+} (y)}{W} \left[ \int^{y} dx ~  (\psi^{(1)}_{\pm} (x,z) T^{(1)}(x,z) + T^{(2)}(x,z)\psi^{(0)}_{\pm}(x))  \psi^{(0)}_{-} (x) \right],
\ea
\ee
where $W$ is a Wronskian of the zeroth order solutions and for \eqref{2BP} equals $\alpha$. 

Notice that the zeroth order solutions have a trivial monodromy along contour $\Gamma$, enclosing points $0$ and $z$, hence the  If we move the first order solutions \eqref{f1s2} along  $\Gamma$, the monodromy matrix up to the first order in the light dimensions has the form (see Appendix D.1 of \cite{Fitzpatrick:2014vua} for details)
\be
\label{mon4}
M_{ab}(z) =\begin{pmatrix}
 1& I^{(2)}_{+-}(z)\\
  I^{(2)}_{-+} (z)& 1
\end{pmatrix} + \cO(\epsilon^2_1, \epsilon^2_2), 
\ee
where the monodromy integrals (the superscript here shows that these are integrals for the system of two heavy operators) pick a contribution from residues at points $0$ and $z$
\be
 \frac{\alpha  I^{(2)}_{+-}} {2 \pi i} = \alpha \epsilon_1 +  c^{(1)}_2(z) (1-z) - \epsilon_2 + (1-z)^{\alpha}\left( c_2(z) (1-z) - \epsilon_2(1 + \alpha)\right) , \quad I^{(2)}_{-+} [\alpha] = - I^{(2)}_{+-}[-\alpha].
\ee
The monodromy equations arise from equating the eigenvalues of the monodromy matrices given in \eqref{mon4} and \eqref{TM} (we put  $p=1$, so $\tilde{\epsilon} \equiv \tilde{\epsilon}_1$). Under the assumption $\tilde{\epsilon} \ll \epsilon_H$, this leads to
\be
\label{4f}
I^{(2)}_{+-} I^{(2)}_{-+}  = -4\pi^2 \tilde{\epsilon}^2. 
\ee
\paragraph{The second order of the HL approximation.} 

The monodromy properties of the second-order solutions are determined by contour integrals of the first order solutions 
$\psi^{(1)}$ over the contour $\Gamma$. While these integrals generally exhibit a complex structure (see \cite{Beccaria:2015shq, Bombini:2018jrg} for explicit expressions), we restrict our analysis to $\alpha=\sqrt{1-4\epsilon_H} \rightarrow 1$ \cite{Fitzpatrick:2014vua}, corresponding to $\epsilon_H\ll 1$. To clarify the relationship between this limit and the light classical dimensions $\epsilon_{1,2}$, we adopt the choice  $\epsilon_1 = \epsilon_2 = \epsilon_H \equiv \epsilon \ll 1$. \footnote{This decomposition differs from \cite{Beccaria:2015shq} (eq. 4.6), where the limit $\epsilon_H\ll 1$ is applied to second-order accessory parameters, whereas here it is applied directly to first order solutions.}

In this regime, the first order solutions simplify to
\be
\ba{c}
\dps \psi^{(1)}_{+}(y,z) = \frac{\epsilon \left(-((y (z-2)+z) \log (y))+(y-1) (z-2) \log (y-z)+2 (z-1) \log \left(\frac{z-y}{z-1}\right)\right)}{z}, \\
\dps \psi^{(1)}_{-}(y,z) = \epsilon \left(\frac{1-y}{z-1}+\left(\frac{2 y}{z}-1\right) \log (y)+\frac{(z-2) \log (y-z)-2 (y-1) \log \left(\frac{z-y}{z-1}\right)}{z}+y-3\right). 
\ea
\ee
For the identity block ($\tilde{\epsilon}=0$), the derivation of monodromy equations follows analogously to the first order case in \eqref{4f}. Equating the monodromy matrices of the second-order solutions \eqref{s3} and \eqref{TM} yields and solving for $c^{(2)}_2 (z)$ 
\begin{equation}
\label{so}
c^{(2)}_2 (z) = \frac{4 \epsilon^2 \left((z-2) z + 2 (z-1) \log (1-z)\right)}{(z-1) z^2}.
\end{equation}

\paragraph{Three heavy operators.}

Here we consider the 4-pt classical with three heavy operators of classical dimensions $\epsilon_1, \epsilon_{3}$ and $\epsilon_{4}$, located at points $(0, 1, \infty)$, respectively (the HHHL block) and one light operator (classical dimension $\epsilon_2$) located at $z$, such that 
\be
\epsilon_2 \ll \epsilon_{1,3,4}. 
\ee
The narration repeats what was written above, so we just list results. The components of stress tensor read
\be
\label{T3M}
T^{(0)} (y) = \frac{\epsilon_1}{y^2} + \frac{\epsilon_{H}}{(1-y)^2} + \frac{\epsilon_1}{y(1-y)}\;, 
\ee
\be
\ba{c}
\dps T^{(1)}(y,z)= c^{(1)}_2(z)\,\frac{(1-z)z}{y(1-y)(y-z)} + \frac{\epsilon_2}{(y-z)^2}+\frac{\epsilon_2}{y(1-y)}\;, \\
\dps T^{(2)}(y, z) = c^{(2)}_2(z) \frac{ z (z - 1)}{y(y - z) (y - 1)}.
\ea
\ee
where we assume $\epsilon_{3}=\epsilon_{4}\equiv \epsilon_H$ for simplicity, but the general case of three different operators can be considered as well. The zeroth order solutions of \eqref{BPZ012} with the stress tensor \eqref{T3M} for this case are
\be
\ba{c}
\label{sol3}
\psi^{(0)}_{\pm}(y) = \dps (1-y)^{\frac{1+\alpha}{2}} y^{\frac{1\pm\beta}{2}} ~ _2F_1 \left(\frac{1\pm \beta}{2},\frac{1\pm \beta}{2}+\alpha, 1\pm \beta|y\right),  \\
\\
\alpha = \sqrt{1 - 4 \epsilon_H}, \qquad \beta = \sqrt{1 - 4 \epsilon_1}. 
\ea
\ee
The monodromy matrix of  the zeroth order solutions \eqref{sol3} is $diag(-\exp[i \pi \beta],-\exp[i \pi \beta])$. Comparing it with \eqref{TM}, we have a constrain $\tilde{\epsilon} = \epsilon_1$ \cite{Alkalaev:2019zhs}. Then, one can compute the monodromy matrix of the first order solutions with help of \eqref{sol3} and \eqref{f1s2}. Notice that due solutions \eqref{sol3} have a branch point at $y=0$, the monodromy of the first order solutions (which is determined by $\int_{\Gamma} dy~ \psi_{\pm}^{(0)} (y)T^{(1)}(y,z) \psi_{\pm}^{(0)} (y)$) is governed only by the pole at $y=z$. It was shown \cite{Alkalaev:2019zhs} that, in contrast to the HHLL block, the monodromy equation is only governed by $I^{(3)}_{++}$ 
\be
\label{4PC_3f}
 I^{(3)}_{++} =0, \qquad I^{(3)}_{++} =\frac{2 i \pi^2}{\sin \pi \beta} \left( c^{(1)}_2(z) \psi_{+} (z) \psi_{-}(z) + \frac{d}{dz} \left( \psi_{+} (z) \psi_{-}(z) \right)\right)\;,
\ee
where $\psi_{\pm}(z)$ are the zeroth order solutions \eqref{sol3} at point $y=z$. Here, we do not consider the second order in the HL approximation because for the simple case $\alpha=1$ even the zeroth order solutions contain logarithms, leading the complicated form of the first order solutions and their monodromy. 

\subsection{Dual description}

\label{sec:HL_A}

In this subsection we recall the Steiner tree problem on the Poincare disk model in the context of AdS/CFT. For the detailed review,  see \cite{Pavlov:2021lca, Alkalaev:2018nik}.    

\paragraph{The Poincare disk and distances.} Let $\mathbb{D} =\{ z \in \mathbb{C}: |z|<1 \} $ denote the Poincare disk with the metric 
\be
ds^2 = \frac{4 dz d\bar{z}}{(1- \bar{z} z)^2}, 
\ee
 and the boundary of the disk ($\partial \mathbb{D}$) is given by the unit circle $|z|=1$. It is convenient to parametrize the disk's interior by $z=t \exp[i \phi], ~ t \in [0,1), ~ \phi \in [0, 2\pi)$, and the boundary $\partial \mathbb{D} = \{ \exp[i w], w \in [0, 2\pi)\}$. The distance between two points $z_1$ and $z_2$ from $\mathbb{D}$ reads 
\be
\label{L}
L(z_1, z_2) = \log \frac{1+u}{1-u}, \qquad u = \frac{|z_1 - z_2|}{|1- \bar z_1 z_2|}\;.
\ee
If one or two points belongs to the boundary\footnote{Despite the fact that the distances \eqref{OB} and \eqref{TB} are divergent, we omit infinite constants, depending on the regulator $\varepsilon$ focusing on the first terms in these formulas, referring to them as regularized lengths. }, the length \eqref{L} becomes infinite \cite{Alkalaev:2018nik}. For one boundary point  $z_1 = \exp[i w_1]$, the length reads
\be
\label{OB}
 L(t, \phi, w_1) = \log \frac{ 2 \left(t^2-2 t \cos(\phi-w_1 )+1\right)}{1-t^2} - \log \varepsilon, \qquad  \varepsilon   \to 0,  
\ee
and for two boundary points $z_{1,2} = \exp[i w_{1,2}]$, is given by 
\be
\label{TB}
 L(w_1, w_2) = 2\log \sin \frac{w_2 - w_1}{2} - 2\log \varepsilon, \qquad  \varepsilon   \to 0. 
\ee

\paragraph{The Steiner tree problem.} The Steiner tree problem is the following: given $N$ points (in our case, they belong to $\mathbb{D}$), find a connected tree of minimal total (weighted) length with such endpoints. More precisely, the tree is characterized by a set of Steiner-Fermat points, linked to each other (by inner edges) and initial points (by outer edges) such that the weighted length 
\be
\label{minimum}
 L_{N} = \sum_{\{\text{outer edges}\}} \epsilon_i L_i + \sum_{ \{\text{inner edges}\}} \tilde \epsilon_j \tilde L_j \;,
\ee becomes minimal. In \eqref{minimum}$, \epsilon_i$ and $\tilde{\epsilon}_j$ denote weights of outer and inner edges, respectively. 

The solution of the problem (the Steiner tree) consists of $(N-2)$ Steiner-Fermat points, which have to be trivalent vertices such that the angles between edges with weights $\epsilon_a, \epsilon_b, \epsilon_c$ intersecting at any  Steiner-Fermat point are determined by \be\label{cooos}
       \cos \gamma_{ac} = \frac{-\epsilon^2_c - \epsilon^2_b + \epsilon^2_a}{2 \epsilon_a \epsilon_c}\;,\quad
       \cos \gamma_{bc} = \frac{-\epsilon^2_c + \epsilon^2_b - \epsilon^2_a}{2 \epsilon_c \epsilon_b}\;,\quad
        \cos \gamma_{ab} = \frac{\epsilon^2_c - \epsilon^2_b - \epsilon^2_a}{2 \epsilon_a \epsilon_b} \;.
\ee
The conditions \eqref{cooos} determine the positions of Steiner-Fermat points, having which we find the weighted length \eqref{minimum} of the Steiner tree. Notice that weights in \eqref{cooos} are restricted by the triangle inequalities
\be
\label{triangle}
\epsilon_a + \epsilon_b \geq \epsilon_c\;,
\qquad
\epsilon_a + \epsilon_c \geq \epsilon_b\;,
\qquad
\epsilon_b + \epsilon_c \geq \epsilon_a\;.
\ee

\paragraph{AdS/CFT correspondence.}

One application of Steiner trees in the Poincaré disk appears in \cite{Alkalaev:2018nik}, where the lengths of holographic Steiner trees with conical defects compute $H^2L^{n-2}$ classical blocks:
\be
\label{fS}
f_{n}(z|\epsilon_h,\epsilon, \tilde\epsilon) = -L_{n-1}(\alpha w|\epsilon, \tilde\epsilon) + i\sum_{k=1}^{n-2}\epsilon_k w_k;,
\ee
with $z(w) = 1 - e^{-i w}$ and $\alpha = \sqrt{1 - 4 \epsilon_H}$. Here $L_{n-1}(\alpha w|\epsilon, \tilde\epsilon)$ is the length of a Steiner tree embedded in the conical defect geometry $\mathbb{D}_{\alpha} \equiv \{ (t, \phi) \mid t \in [0,1], \phi \in [0, 2\pi \alpha) \}$. The tree has $(n-2)$ endpoints on the boundary and one endpoint at the origin ($t=0$). The angle deficit $\alpha$ derives from the classical dimension $\epsilon_H$ of heavy operators
ana the classical block $f_{n}(z|\epsilon_H,\epsilon, \tilde\epsilon)$ is obtained by identifying classical dimensions with Steiner tree weights and applying the coordinate transformation $z(w) = 1 - e^{-i w}$.

In what follows, we focus on the particular example of the tree dual to the HHLL block. 

\section{ODEs for HL blocks from the monodromy method}
\label{sec:MODE}
In this Section, we apply the monodromy method to derive ODEs for the non-identity HHLL block, mainly focusing on the case  $\epsilon_1 = \epsilon_2$. Then, we generalize the derivation above to HL blocks with three heavy operators.

\subsection{First order HL blocks}

\label{sec:4pt2H}

\paragraph{HHLL blocks $\epsilon_1 = \epsilon_2.$}

Recall the monodromy system for the block \eqref{4f}
\be
\label{IE1}
I^{(2)}_{+-} I^{(2)}_{-+}  = -4\pi^2 \tilde{\epsilon}^2,
\ee
where 
\be
I^{(2)}_{+-} = \frac{2 \pi i \epsilon_2}{\alpha} \left[\alpha + 2 C(z) (1-z) - 1 + (1-z)^{\alpha}\left(2 C(z) (1-z) - 1 - \alpha\right)\right], ~~  C(z) = \frac{c^{(1)}_2(z)}{2\epsilon_2},
\ee
and $C(z)$ stands for the "dimensionless" accessory parameter which is our main object in this Section. Taking a derivative of the equation \eqref{IE1} with respect to $z$, we get 
\be
\label{4D}
\left( 2 \alpha \left((1-z)^{\alpha}+1\right) C(z)-\left((1-z)^{\alpha}-1\right) \left(\alpha^2+4 (1-z) C'(z)\right) \right) = 0.
\ee
By excluding $(1-z)^{\alpha}$ from \eqref{IE1} and \eqref{4D}, one finds the following ODE (parameterizing by $\kappa$) for $C(z)$
\be
\label{4FE}
(C'(z) -C^2(z) -T_2(z))^2= \kappa^{2}\left(\frac{\alpha^2 ((1-z)C(z) - \half)^2}{4} -\left( \frac{\alpha^2 - 4(1-z)C(z)}{4(1-z)^2} +  C'(z)\right)^2\right),
\ee
where 
\be
\label{T2}
T_2(z) = \frac{1 - \alpha^2}{4(1-z)^2} = \frac{\epsilon_H}{(1-z)^2}, \qquad \kappa \equiv \frac{\tilde \epsilon}{2\epsilon_2}. 
\ee There are two simple cases, when \eqref{4FE} reduces to non-homogeneous Riccati equations for $C(z)$ 
\begin{subequations}\label{sc}
\be
\label{4PC}
\kappa = 0: \quad C'(z) = C^2(z) + T_2(z),
\ee
\be
\kappa = 1: \dps \quad C'(z) = \half C^2(z) +\frac{ C(z)}{2(1-z)}+ \frac{T_2(z)}{2}. 
\ee
\end{subequations}
The first one corresponds to the identity block $\tilde \epsilon=0$. Using that $C(z) = \frac{1}{2\epsilon_2} \frac{d  f(z|\epsilon_2, \epsilon_H)}{dz}$, we rewrite  \eqref{4PC} as 
\be
\label{4vPC}
\frac{1}{2\epsilon_2}\frac{d^2 f(z|\epsilon_2, \epsilon_H)}{dz^2} = \frac{\epsilon_H}{(1-z)^2} + \frac{1}{4 \epsilon_2^2} \left(\frac{d  f(z|\epsilon_2, \epsilon_H)}{d z} \right)^2,
\ee
which was originally derived in \cite{Fitzpatrick:2015foa} by the different approach. The second case saturates one of conditions \eqref{triangle}, which leads to simplification. The solutions are determined by one arbitrary constant,  which can be fixed by the asymptotic behavior of $C(z) \rightarrow \frac{\kappa-1}{z}$ at $z \rightarrow 0$. 

The technique described above can be applied to a more general case of $4$-pt blocks with $\epsilon_1 \neq \epsilon_2$. Despite the fact that the general ODE is cumbersome, for cases $\epsilon_1 = 2 \epsilon_2 = 2 \tilde{\epsilon}$ and $\epsilon_2 = \tilde{\epsilon} = 2 \epsilon_1$ it reduces to Riccati equation.  Another example, which plays an important role for a future discussion, requires $\tilde \epsilon = \epsilon_1 \ll \epsilon_2$. For this case, we have 
\be
\label{Lim3H}
 \tilde{C}'(z) =\tilde{C}^2-\frac{(1+\alpha) \tilde{C}(z)}{1-z}+ \frac{1+\alpha}{(1-z)^2}, \qquad \tilde{C}(z) = \frac{c^{(1)}_2(z)}{\epsilon_2}, 
\ee
where we change normalization for future needs.

\paragraph{ODE for the HHHL block.} 
The simplest case beyond HL blocks with two heavy operators is the HHHL block. The monodromy equation for the block reads \eqref{4PC_3f}
\be
\label{4PC_3}
 I^{(3)}_{++} =0, \quad I^{(3)}_{++} =\frac{2 i \pi^2 \epsilon_2}{\sin \pi \beta} \left( C(z) \psi_{+} (z) \psi_{-}(z) + \frac{d}{dz} \left( \psi_{+} (z) \psi_{-}(z) \right)\right)\;, \quad C(z) = \frac{c^{(1)}_2(z)}{\epsilon_2},
\ee
where $\psi_{\pm}(z)$ are the zeroth order solutions \eqref{sol3}. Notice that we change normalization of the HHHL block (previously, HHLL blocks were divided by $2\epsilon_2$). In contrast to HL blocks with two heavy operators, we take derivative of \eqref{4PC_3} twice, and simplifying using hypergeometric function identities yields 
\be
\label{E3H}
C''(z) = -C^3(z)+  3 C(z) C'(z)- 4 T_3(z) C(z) + 2 T_3'(z),
\ee
where 
\be
\label{T3}
T_3(z) = \frac{\epsilon_1}{z^2} + \frac{\epsilon_H}{(1-z)^2} + \frac{\epsilon_1}{z(1-z)}. 
\ee 

\paragraph{The HHLL block as a limit of the HHHL block.} As it was pointed out in \cite{Alkalaev:2019zhs}\footnote{See Appendix in \cite{Alkalaev:2019zhs} for details concerning series expansion of classical $4$-pt blocks.  There are also works related to various limits of large-$c$ blocks \cite{Anous:2020vtw, Alkalaev:2024knk}. }, in the limit $\epsilon_1 \rightarrow0$ (or $\beta =  1 - 2\epsilon_1 + \cO(\epsilon_1^2)$) we get the $4$-pt block with $\tilde \epsilon = \epsilon_1$ for the HHHL block. An important detail is that by taking this limit we do not keep ratio $\epsilon_1/\epsilon_2$ to be finite. Hence, the resulting $4$-pt block is a HHLL block with $\epsilon_1=\epsilon_p \ll \epsilon_2$ and we should compare the limit of the equation \eqref{E3H} to \eqref{Lim3H}. 

Foremost, obviously $T_3(z) \rightarrow T_2(z)$ at $\epsilon_1 \rightarrow 0$. Then, it is easy to see that 
\be
\label{LP}
 \left(\frac{d}{dz} - \tilde{C}(z) \right) \eqref{Lim3H} = 0 : \quad  \tilde{C}''(z) = - \tilde{C}^3(z)+  3  \tilde{C}(z)  \tilde{C}'(z)- 4 T_2(z) \tilde{C}(z)  + 2 T_2'(z),
\ee
which is explicitly \eqref{E3H} for $\epsilon_1=0$. So, it means that there is a higher (third) order equation for $\tilde{C}(z)$ which governs the limiting case of \eqref{E3H}. Conversely, putting $\epsilon_1=0$ we can reduce \eqref{E3H} to the first order equation \eqref{Lim3H}. 

\subsection{Second order HL blocks}
\label{sec:SO}
We now analyze second-order corrections to the HHLL identity block \eqref{so} in the second order (for $\alpha=1$). Similar to the first order, one can show that \eqref{so} satisfies
\be
\frac{d c^{(2)}_2(z)}{d z} + \frac{2}{z} c^{(2)}_2(z) = \frac{4 \epsilon}{(1-z)^2},
\ee
which is the Bernoulli differential equation. To extend the results found in the first order, one can define 
\be
D(z) = \frac{c_2^{(1)}(z)+c^{(2)}_2 (z)}{2 \epsilon},  
\ee
we obtain the following Riccati equation (up to  $\epsilon^2$)
\be
\label{soq}
\quad D'(z) = D^2(z) + \frac{2\epsilon}{(1-z)^2}. 
\ee
The equation maintains the same structure as \eqref{4PC}, but differs in the classical dimension appearing on the right-hand side ($2\epsilon$ instead of $\epsilon_H$). For $\alpha=1$, the contribution from $c^{(1)}_2$ vanishes, so the right-hand side is determined exclusively by  $ c^{(2)}_2$. When we impose $\epsilon_H \equiv \epsilon \ll 1$ in \eqref{4PC}, we recover \eqref{soq} to first order in $\epsilon$. This agreement demonstrates that \eqref{4PC} remains valid to second order for this configuration.

The analogous result discussed in \cite{Fitzpatrick:2016mtp}, where it was derived from the dual description in terms of Wilson lines. The Ricatti equation appeared as the equation of motion for the boundary Hamiltonian (see Section 4 in \cite{Fitzpatrick:2016mtp}). It was checked that the  HHLL block satisfies these equation in several first order in $1/c$. For the general $\alpha$ the second order accessory parameter does not satisfy any Ricatti equation, but there are examples of such equations for certain values of $\alpha$, see Appendix B.2 in \cite{Beccaria:2015shq}.

\subsection{Exponentiated classical blocks and BPZ equations}
In this subsection we analyze the non-linear equations for the HHLL and HHHL blocks. More precisely, we introduce exponentiated classical blocks as 
\be
\label{expb}
v^{(1)}(z) = \exp \left[-\frac{f^{(1)}(z| \epsilon, \tilde{\epsilon})}{\epsilon_2} \right]. 
\ee Such blocks represent $\tilde{F}_{4}(z| \Delta_2, \Delta_4)^{- \Delta_2}$, where $\tilde{F}_{4}$ is a conformal block, corresponding to   $f^{(1)}(z| \epsilon, \tilde{\epsilon})$.  It can be shown that these exponentiated blocks satisfy the BPZ equation for the singular vector $V_{(1,3)}$. 

To prove it, we start from the Ricatti equation for the identity HHLL block \eqref{4PC}. It is known that any Riccati equation can be converted to a second order ODE \cite{nehari}. Indeed, for $u(z)$ satisfies the general Riccati equation 
\be
u'(z)= r(z) u^2(z) + p(z)u(z)+q(z), 
\ee
one considers a substitution $u(z) = \dps -\frac{\psi'(z)}{r(z)\psi(z)}$, which yields 
\be
r(z) \psi''(z) - (r'(z) +r(z) p(z)) \psi'(z) + r^2(z) q(z) \psi(z) = 0. 
\ee
In the simplest case of the identity block \eqref{4PC} $r(z)=1,~ p(z)=0,~ q(z) = T_2(z)$, so we end up with 
\be
\label{QB}
u''(z) + T_2(z)  u(z) = 0, \qquad u(z) = \exp\left[-\int^{z} C(w)dw \right].
\ee
We see that \eqref{QB} coincides with the BPZ equation (associated with the singular vector $V_{(1,2)}$) in the zeroth order with the stress tensor \eqref{T2}. As well as for two heavy operators, by the similar procedure we convert \eqref{E3H} to  a third order ODE
\be
\label{V13}
\quad  u'''(z) + 4 T_3(z) u'(z) + 2T_3'(z) u(z) = 0, \qquad u(z) = \exp\left[-\int^{z} C(w)dw \right],
\ee
which is the BPZ (associated with the singular vector $V_{(1,3)}$) equation with the stress tensor \eqref{T3} (see also \cite{Chen:2016cms}, where the BPZ equation above arises in discussion of $1/c$ corrections.)

Since for three heavy operators $u(z)$ in \eqref{V13} coincides with the \eqref{expb}, we see that the exponentiated block satisfies the BPZ equation for the singular vector $V_{(1,3)}$ with stress tensor \eqref{T3}. For the case of two heavy operators, \eqref{expb} is equal to $u^2(z)$ in \eqref{QB} and satisfies the BPZ equation for the singular vector $V_{(1,3)}$ as well (since any product of two solutions of the BPZ equations for the singular vector $V_{(1,2)}$ solves the BPZ equations for the singular vector $V_{(1,3)}$ \cite{Zamolodchikov:2003yb}). 

The appearance of the BPZ equations can be explained as follows. For three heavy operators, it follows directly from \eqref{4PC_3} that the dimensionless accessory parameter $C(z)$ is the logarithmic derivative of the product of two solutions to the BPZ equation with stress tensor \eqref{T3}. As noted earlier, this product satisfies the BPZ equation for the singular vector $V_{(1,3)}$. For two heavy operators, the connection is less straightforward: monodromy equations show that $C(z)$ is the logarithmic derivative of a fractional linear function of solutions to the BPZ equation with stress tensor \eqref{T2}. Since $v^{(1)}(z) \sim \exp \left[ - 2 \int^{z} dy ~ C(y)\right]$, it satisfies the BPZ equation for the singular vector $V_{(1,3)}$.

To conclude, recall that the conformal block can be straightforwardly constructed once the equation for the exponentiated block \eqref{expb} is known a simpler and more tractable object. Remarkably, an analogous situation arises in four-dimensional conformal field theory, where conformal blocks satisfy higher-dimensional generalizations of BPZ-like differential equations \cite{Huang:2023ikg}. It would be interesting to try to see the structures associated with non-identity blocks in d-dimensional blocks. 

\section{ODEs for the HL blocks from the dual description}
\label{sec:HODE}
In this Section we derive ODEs for the HHLL blocks with $\epsilon_1 = \epsilon_2$ using their dual description given by  \eqref{fS}. 

\begin{figure}[H]
\centering
\begin{tikzpicture}[scale=2.]
 \tkzDefPoint(6,0){OO}
  \tkzDefPoint(7,0){B}
  \tkzDrawCircle[line width  = 0.3](OO,B)

  \tkzDefPoint(2,0){OOO}

  \tkzDefPoint(6.5,-0.855){w1}
  \tkzDefPoint(5.5,-0.855){01}
  \tkzDefPoint(5.92,-0.4){x1}

  \tkzDefPoint(6,-1.158){a}

 \tkzDefPoint(6.777,-1.615){pp_1}
 \tkzDefPoint(5.618,-1.382){pp_2}


 \tkzDrawArc[rotate,color=red,line width  = 1.0](pp_2,w1)(42)
  \tkzDrawArc[rotate,color=green,line width  = 1.0](pp_1,01)(-24)
  \draw[blue,thick](6,0) -- (5.92,-0.4);

\tkzDefPoint(6,-1.15){dr}

\tkzDrawArc[dashed, rotate,color=black,line width  = 0.3](dr,w1)(122)
\draw[dashed,color = black, line width  = 0.3] (6, 0) -- (6.5,-0.855);
\draw[dashed,color = black, line width  = 0.3] (6, 0) -- (5.5,-0.855);

 \tkzDefPoint(2.8,-0.6){z_{3}}
  \tkzDefPoint(1.055,-0.29){z_{1}}
  \tkzDefPoint(2.1,-0.991){z_{2}}
  \tkzDefPoint (1.345, -1.31){c_c_1}
  \tkzDefPoint (3.1, -0.73){c_c_2}
  \tkzDefPoint (2.4,-1.13){c_c_3}

 \tkzDefPoint (1.283, -1.075){c_o_1}
  \tkzDefPoint (2.537, -0.956){c_o_2}
  \tkzDefPoint (1.62,-2.17){c_o_3}

  \tkzDefPoint(2.08,-0.55){xx1}

 \tkzDrawPoints[color=black,fill=white,size=2](x1,w1,01)
 \tkzDrawPoints[color=black,fill=black,size=2](OO)

  \draw (5.45, -1.04) node {$w_1$};
  \draw (6.55, -1.04) node {$w_2$};

\end{tikzpicture}
\caption{A Steiner tree dual to a non-identity HHLL block. The tree's segments are depicted in different colours correspond to different weights. A triangle with corners at boundary endpoints and at the center of the disk is shown in dashed. }
\label{Steiner}
\end{figure}
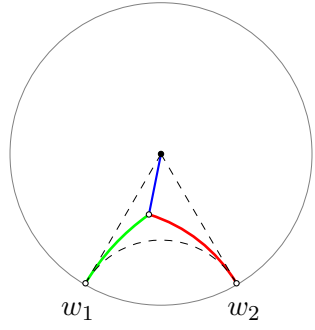

The corresponding Steiner tree (Fig.~\ref{Steiner}) consists of three segments: two boundary segments ($Y_1$ and $Y_2$) connecting the endpoints to the Steiner-Fermat point and one bulk-to-bulk segment of length $X$. The (regularized) lengths of the segments are related by \cite{Pavlov:2021lca}
\be
\ba{c}
\label{4Steiner}
  2 \exp[-Y_1]=  (\cosh X - \sinh X \cos  \gamma_{13})\;,  
  \\
    2 \exp[-Y_2]=  (\cosh X - \sinh X \cos  \gamma_{12})\;, 
    \\
    Y_1 + Y_2 = 2 \log \sin \frac{w}{2} + \dps 2 \log \sin \frac{\gamma_{12}}{2}, 
\ea
\ee
where angles between edges are given by \eqref{cooos}. For the $\epsilon_1 = \epsilon_2$, the tree exhibits symmetry ($\cos\gamma_{13} = \cos\gamma_{12}$, hence $Y_1 = Y_2$). The weighted length $L(w) \equiv L_3(w|\epsilon_2, \tilde{\epsilon})$ is then given by
\be
\label{WL}
L = 2 \epsilon_2 Y + \tilde{\epsilon} X.
\ee  

\paragraph{A tree with $\tilde{\epsilon} =0$.} First, we analyze the case $\tilde{\epsilon}=0$. The weighted length takes the form \eqref{WL}:
\be
\label{WW}
L = 2 \epsilon_2 \log \frac{1 + t^2 - 2 t \cos \left(\frac{w}{2} \right)}{1- t^2},
\ee
which consists of two equal bulk-to-boundary segments \eqref{OB}. The symmetry of this tree requires $\phi = w/2$. Taking the total derivative of \eqref{WW} twice with respect to $w$ and using the minimization condition for $t(w)$ (the radial coordinate of the Steiner-Fermat point) yields
\be
\label{HawkingfromCatalanfromSteiner}
\dot{Y}^2 - \ddot{Y} = \frac{1}{4},
\ee
where dots denote $d/dw$. 
This equation can be adapted for the HHLL block using \eqref{fS}:
\be
\frac{f_4(z(w)|\epsilon_H)}{2 \epsilon_2} = - Y(\alpha w) + \frac{i w }{2}, \qquad w = -i \log (1-z),
\ee
yielding \eqref{4vPC} \cite{Fitzpatrick:2015foa}.

\paragraph{A tree with $\tilde{\epsilon} \neq 0$.} Deriving an ODE for the length itself is more challenging here. Instead, we establish differential relations between segments $X$ and $Y$ in \eqref{WL}. The weighted length becomes
\be
L = 2 \epsilon_2 \log \frac{1 + t^2 - 2 t \cos \left(\frac{w}{2} \right)}{1- t^2} + \tilde{\epsilon} \log \frac{1+t}{1-t}.
\ee
Applying the same minimization logic as above gives
\be
\label{mf}
\frac{\ddot{X}}{\dot{X}} = \dot{Y}.
\ee
A second relation follows from \eqref{4Steiner}. For the symmetric case,
\be
2 \exp[-Y] = \cosh X - \sinh X \cos \gamma_{13},
\ee
and differentiating twice with respect to $w$ produces
\be
\label{ntbi}
-\dot{Y}^2 + \ddot{Y} = \frac{\ddot{X}}{\dot{X}} \dot{Y} - \dot{X}^2.
\ee

To derive an ODE for $l(w) \equiv L(w)/(2 \epsilon_2)$, we count equations and variables: three equations (\eqref{HawkingfromCatalanfromSteiner}, \eqref{ntbi}, \eqref{mf}) govern four variables (first and second derivatives of $X$ and $Y$). This reduces to one equation for $l(w)$:
\be
\label{MI2}
\left(\dot{l}^2 - \ddot{l} + \frac{1}{4}\right)^2 = \kappa^2 \left( \left(\frac{1}{4} - \ddot{l}\right)^2 + \frac{\dot{l}^2}{4} \right), \qquad \kappa = \frac{\tilde{\epsilon}}{2 \epsilon_2}.
\ee
The relation
\be
C(z) = \frac{i}{1-z} \left(\frac{i}{2} - \dot{l} (\alpha w)\right)
\ee
connects \eqref{MI2} and \eqref{4FE}. Substituting $\dot{l}(w) = k(w)$ yields a first order nonlinear ODE for $k(w)$ that admits separation of variables, as discussed in Section \bref{sec:MODE}.

\section{Conclusion}
\label{sec:conclusion}

In this paper, we derived ODEs for the HL classical blocks. The accessory parameters for HHLL and HHHL blocks are subject to first and second order non-linear differential equations, respectively. We presented the ODE for the HHLL block with $\epsilon_1 = \epsilon_2$ and discussed special cases leading to Riccati equations. Conversely, the HHHL block satisfies the second-order ODE \eqref{E3H}, a generalization of the Riccati equation. We also analyzed the limiting procedure connecting HHHL and HHLL blocks. To extend the results, we considered one particular case of the identity HHLL in the second order of the HL approximation in Section \bref{sec:SO}. The connection between the ODEs for classical blocks and the BPZ equations was established. Also, we derived the equations for the HHLL blocks from the dual description.  

It is worth mentioning several future directions. The first one is to combine the Painlevé VI approach to classical blocks \cite{Litvinov:2013sxa} (which relies  on the classical limit only) with the HL approximation to address higher orders. The second one is generalizing the above analysis to identity \cite{deBoer:2014sna} and semi-degenerate \cite{Fateev:2007ab, Fateev:2008bm, Belavin:2016qaa, Belavin:2016wlo} 4-pt $W_3$ blocks, as their ODEs remain unknown despite recent progress \cite{Karlsson:2021mgg}. 

\paragraph{Acknowledgments.}I am grateful to Aleksandr Artemev, Vladimir Belavin, Igor Chaban, Alexey Litvinov, Andrei Marshakov and Danil Zherikhov for numerous discussions. The work was supported by the Foundation for the Advancement of Theoretical Physics and Mathematics “BASIS”.

\providecommand{\href}[2]{#2}\begingroup\raggedright\endgroup

\end{document}